\documentclass[12pt]{iopart}
\usepackage{color,iopams,epsfig}

\begin{document}

\title{Generalized q-Deformed Symplectic sp(4) Algebra for 
Multi-shell Applications}
\author{K. D. Sviratcheva$^1$, A. I. Georgieva$^{1,2}$ and J. P. Draayer$^1$}
\address{$^1${\it Louisiana State University, Department of Physics 
and Astronomy,
Baton Rouge, Louisiana, 70803-4001 USA}
\\
$^2${\it Institute of Nuclear Research and Nuclear Energy,
Bulgarian Academy of Sciences, Sofia 1784, Bulgaria}}

\begin{abstract}
A multi-shell generalization of a fermion representation of the 
$q$-deformed compact symplectic 
$sp_q(4)$ algebra is introduced. An analytic form for the action of two or
more generators of the $Sp_q(4)$ symmetry on the basis states is determined
and the result used to derive formulae for the overlap between number
preserving states as well as for matrix elements of a model Hamiltonian.
A second-order  operator in
the generators of $Sp_q(4)$ is identified that is diagonal in the basis
set and that reduces to the Casimir invariant of the  $sp(4)$ algebra in
the non-deformed limit of the theory. The results can be used in nuclear
structure applications to calculate $\beta$-decay transition
probabilities and  to provide for
a description of pairing and higher-order interactions in systems with nucleons
occupying more than a single-$j$ orbital.
\end{abstract}

\maketitle

\section{Introduction}

The symplectic $sp(4)$ algebra, which is isomorphic to $so(5)$ \cite{Hecht,
Ginocchio, GoLiSp}, has been successfully used for a description of pairing
correlations when two types of particles are taken into account
\cite{EngelLangankeVogel,Dobes}. The algebra can be generalized to 
multiple levels
\cite{Kerman,Ginocchio,TalmiSimpleModels} so the particles can occupy
more than a single orbit. When applied to nuclear structure, this
generalization of the $sp(4)$ algebra makes a study of nuclei with mass
numbers $56<A<100$ possible.  

An additional degree of freedom can be introduced through a $q$-deformation of
the classical $sp(4)$ Lie algebra \cite{Jimbo,Drinfeld}. While this
preserves the  underlying symmetry,
it introduces non-linear terms into the theory. 
In contrast
with the usual formulation of $q$-deformation for the symplectic $sp(4)$
algebra and its $su(2)$ subalgebras that is normally used in 
mathematical studies
\cite{hay,liNoRu,aach} and in nuclear physics applications
\cite{Sharma,SharmaSh,Berej}, we have discovered a new formulation that depends
upon the dimensionality of the underlying space \cite{fermRealSp4}. 
Because of this
dependence, a generalization of the $q$-deformed symplectic $sp_q(4)$ 
algebra to a
multi-orbit case is an interesting exercise that introduces new 
elements into the
theory.

To apply the $q$-deformed theory (single-level or multiple-orbit) to 
real physical
systems, the action of the generators of $Sp_q(4)$ on the basis states must be
known. We have derived an analytical form for accomplishing this. The 
results can
be used to build a $q$-deformed analog of the second-order Casimir 
invariant of the
$sp(4)$ algebra. The results also provide for  analytical matrix 
elements of a model
interaction and therefore for an exact solution of the corresponding 
Hamiltonian
problem. The results are what is needed for nuclear structure 
applications and an
investigation into the physical significance of $q$-deformation.

\section{Generalized sp(4) algebra and its q-deformed extension}

The multi-shell generalization of the fermion realization of $sp_{(q)}(4)$
follows the single-$j$ construction of the algebra \cite 
{fermRealSp4}. The $sp(4)$
algebra, which is isomorphic to $so(5),$ is realized in terms of creation and
annihilation fermion operators $c_{j,m,\sigma }^{\dagger }$ and 
$c_{j,m,\sigma }$,
which describe a particle of type $ \sigma $ ($=\pm 1$ for protons/neutrons) in 
a
state of total angular momentum $j$ (half-integer) with a third projection $m$
($-j\leq m\leq j$).  For a given $\sigma ,$ the dimension of the 
fermion space is
$2\Omega =\sum_{j}2\Omega _{j}=\sum_{j}(2j+1)$, where the sum $\sum_{j}$
is over all orbits that are considered to be active.

The deformation of the $sp_{q}(4)$ algebra is introduced in terms of 
$q$-deformed
creation and annihilation operators $\alpha _{j,m,\sigma }^{\dagger }$ and
$\alpha _{j,m,\sigma },$ $(\alpha _{j,m,\sigma }^{\dagger })^{*}=\alpha
_{j,m,\sigma }$, where $\alpha _{j,m,\sigma }^{(\dagger )}\rightarrow
c_{j,m,\sigma }^{(\dagger )}$ in the limit $q\rightarrow 1.$ The deformed
single-particle operators are defined through their anticommutation relation
for every $j$, $\sigma $, and $m$ in a similar way as for the 
single-level problem
\cite{fermRealSp4}
\begin{equation}
\begin{array}{ll}
\{\alpha _{j,\sigma ,m},\alpha _{j,\sigma ,m^{\prime }}^{\dagger }\}_{\pm
1}=q^{\pm \frac{N_{\sigma }}{2\Omega }}\delta _{m,m^{\prime }}, & \{\alpha
_{j,\sigma ,m},\alpha _{j^{\prime },\sigma ^{\prime },m^{\prime }}^{\dagger
}\}=0,\sigma \neq \sigma ^{\prime },j\neq j^{\prime }, \\
\{\alpha _{j,\sigma ,m}^{\dagger },\alpha _{j^{\prime },\sigma ^{\prime
},m^{\prime }}^{\dagger }\}=0, & \{\alpha _{j,\sigma ,m},\alpha _{j^{\prime
},\sigma ^{\prime },m^{\prime }}\}=0,
\end{array}
\label{qfcr}
\end{equation}
where the two Cartan generators $N_{\sigma
}{=}\sum_{j}\sum_{m=-j}^{j}c{{{_{j,m,\sigma }^{\dagger }}}}c{\ _{j,m,\sigma }}$
count the number of particles of each type $\sigma $ and
by definition the
$q$-anticommutator is given as $\left\{ A,B\right\} _k=AB+q^{k}BA.$

In the deformed case a pair of fermions can be created ($F_{0,\pm 1}
\stackrel{q\rightarrow 1}{\rightarrow } A_{0,\pm 1}$) or destroyed 
($G_{0,\pm 1}
\stackrel{q\rightarrow 1}{\rightarrow } B_{0,\pm 1}$) by the
operators:
\begin{equation}
F_{\frac{\sigma +\sigma ^{\prime}}{2}}=\frac{1}{\sqrt{2\Omega (1+\delta
_{\sigma ,\sigma ^{\prime }})}}\sum_{j}\sum_{m=-j}^{j}{{(-1)}^{j-m}}\alpha
_{j,m,\sigma }^{\dagger }\alpha _{j,-m,\sigma ^{\prime }}^{\dagger },
\label{qgGen1}
\end{equation}
\begin{equation}
G_{\frac{\sigma +\sigma ^{\prime}}{2}}=\frac{1}{\sqrt{2\Omega (1+\delta
_{\sigma ,\sigma ^{\prime }})}}\sum_{j}\sum_{m=-j}^{j}{{(-1)}^{j-m}}\alpha {%
_{j,-m,\sigma }}\alpha {_{j,m,\sigma ^{\prime }}},
\label{qgGen2}
\end{equation}
where $F_{0,\pm 1}=(G_{0,\pm 1})
^{\dagger }$. The number preserving Weyl generators are defined as:
\begin{equation}
T_+=\frac{1}{\sqrt{2\Omega }}\sum_{j}\sum_{m=-j}^{j}\alpha {{{\
_{j,m,1}^{\dagger }}}}\alpha {\ _{j,m,-1}},\quad T_-=\frac{1}{%
\sqrt{2\Omega }}\sum_{j}\sum_{m=-j}^{j}\alpha {{{_{j,m,-1}^{\dagger }}}}%
\alpha {_{j,m,1},}  \label{rlGen}
\end{equation}
where ($T_{\pm } \stackrel{q\rightarrow 1}{\rightarrow } \tau _{\pm 
}$). In addition
to the Cartan generators $N_{\pm 1}$ (or their linear combinations 
$N=N_{+1} +N_{-1}$
and $T_0 \equiv \tau _0=(N_{+1}-N_{-1})/2$), the operators 
(\ref{qgGen1})-(\ref{rlGen})
close on the $q$-deformed $sp_q(4)$ algebra and their non-deformed 
counterparts close
on the $sp(4)$ algebra. In physical applications, the  number 
generators $N_{\pm 1}$
along with the total number of particles $N$ and the third projection 
$T _0$ (of the
operator with two other components $T_{\pm }$), represent physical
observables, which are always non-deformed.

In the deformed and non-deformed cases, the generators
(\ref{qgGen1})-(\ref{rlGen}) are related to the corresponding single-level
operators
$X^{(j)}$ as
$X=\,\sum_{j}\frac{\sqrt{\Omega _{j}}}{\sqrt{\Omega }}X^{(j)}$, where 
$X=\{ F,\ G,\ T
\}$ or $X \stackrel{q\rightarrow 1}{=}\{ A,\ B,\ \tau \}$.
In the non-deformed limit, the ten operators
$A_{0,\pm 1}^{(j)}$, $B_{0,\pm 1}^{(j)}$, $\tau _{0, \pm }^{(j)}$ and
$N^{(j)}$ close on the
$sp^{(j)}(4)$ algebra for each $j$-level and
the direct sum holds, $sp(4)=\bigoplus_{j}sp^{(j)}(4)$.

A different situation occurs in the deformed case, where the
single-level generators do not close within the $sp^{(j)}_q(4)$ algebra (e.g.
$[T^{(j)}_{+},T^{(j)}_{-}]\ne [2\frac{{{T}^{(j)}_{0}}}{2\Omega _j}]$)
but rather within the generalized $sp_q(4)$ algebra, since
\[
{\lbrack }T^{(j)}_{+},T^{(j)}_{-}{]}=[2\frac{{{T}_{0}}}{2\Omega }
],[F^{(j)}_{0},G^{(j)}_{0}]=[\frac{N-2\Omega}{2\Omega }],\left[ F^{(j)}_{\pm
1},G^{(j)}_{\pm 1}\right] =\rho _{\pm }[\frac{N_{\pm 1}-\Omega}{\Omega }],
\]
where by definition $[X]_k=\frac{q^{kX}-q^{-kX}}{q^k-q^{-k}}$ and $\rho _{\pm
}=\frac{q^{\pm 1}+q^{\pm \frac{1}{2\Omega }}}{2}$. The rest of the commutation
relations remain within the single-$j$ $q$-deformed algebra, for example
$[T_0^{(j)},T_{\pm }^{(j)}]=\pm T_{\pm }^{(j)}$. However, several of 
these relations,
like $[T_l^{(j)},F_0^{(j)}]_{\frac{[2]}{2}}=\frac{1}{2\sqrt{\Omega
_j}}F_l^{(j)}(q^{\frac{N_{-l}}{2\Omega }}+q^{-\frac{N_{-l}}{2\Omega 
}})$, $l=\pm 1$,
include a multiplicative $q$-factor with a dependence on the averaged 
multi-level
number, $N_{\pm 1}/(2\Omega )$.
This behavior of $sp_q^{(j)}(4)$ can be traced back to the generalized
$q$-deformation (\ref{qfcr}), where the anticommutation relations of 
two fermions on
a single-$j$ level depend on the total number of particles of one 
kind averaged over
the multi-shell space. Another interesting consequence of (\ref{qfcr}) is  the
single-$j$ $q$-deformed quantity
\begin{equation}
\sum_{m}\alpha _{j,\sigma ,m}^{\dagger }\alpha _{j,\sigma ,m}=2\Omega _{j}[
\frac{N_{\sigma }}{2\Omega }],\ \sigma =\pm 1.
\label{qN}
\end{equation}
In the non-deformed limit, the left-hand side of (\ref{qN}) represents the
single-level number operator $N^{(j)}_{\sigma }$, while in the 
deformed extension
the zeroth approximation of (\ref{qN}) gives an even distribution of 
the particles
over the entire multi-level space weighed by the single-$j$ dimension.
In this way, the $q$-deformation for the generalized $sp_q(4)$ 
algebra introduces
probability features at the constituent single-$j$ levels of the theory.

In the deformed case (as in the ``classical'' case), each finite 
representation is
spanned by completely paired states, which are constructed as pairs of fermions
coupled to a total angular momentum and parity
$J^{\pi }=0^{+}$ \cite {KleinMarshalek},
\begin{equation}
\left| n_{1},n_{0},n_{-1}\right) _{q}=\left( F{{{_{1}}}}\right)
^{n_{1}}\left( F{{{_{0}}}}\right) ^{n_{0}}\left( {{{F_{-1}}}}\right)
^{n_{-1}}\left| 0\right\rangle ,  \label{qGencsF}
\end{equation}
where $F_{0,\pm 1}$ are defined in (\ref{qgGen1}) and 
$n_{1},n_{0},n_{-1}$ are the
total number of pairs of each kind,
$(\sigma , \sigma ^{\prime} )=(++),(+-),(--)$,
respectively. The basis states (\ref{qGencsF}) in a multiple-orbit 
space of dimension
$2\Omega $ is a linear combination of the single-level basis states 
which depend on
what pairs occupy which levels
\cite{KermanLawsonMacfarlane,PanDraayer}.

The states (\ref{qGencsF}) are eigenvectors of the total number 
operators $N_{\pm
1}$ with eigenvalues $N_{\pm }$, where $N_{\pm }=2n_{\pm 1}+n_{0}$.
Both $N$ and $\tau _{0}\ (T_0)$ are diagonal in the basis (\ref{qGencsF}) with
eigenvalues $ n=2(n_{1}+n_{-1}+n_{0})$ and $i=n_{1}-n_{-1}$, 
respectively. While
the single-$j$ fermion number operators $N^{(j)}_{\pm 1}$ project onto the
single-level basis, the $q$-deformed analog (\ref{qN}) is diagonal in the basis
(\ref{qGencsF}) with eigenvalue $2\Omega _j [
\frac{2n_{\sigma }+n_{0}}{2\Omega }]$, $\sigma =\pm 1$.

\begin{table}[b]
\caption{Reduction limits of the $sp_{q}(4)$ algebra, $\mu =\{T ,0,\pm\}$.}
\begin{tabular}{|c|c|c|c|c|}
\hline
$u_{q}^{\mu }(2)$ & Generators &\multicolumn{2}{c|}{ Eigenvalues }
& Basis states \\ \cline{3-4}
& & ${\bf C}_{1}(u_{q}^{\mu }(2))$ & ${\bf C}_{2}(su_{q}^{\mu }(2))$ 
& \\ \hline
$u_{q}^{T }(2)$ & $T_{\pm},\ T _0;\ N$ &
$\begin{array}{c}
n= 2n_{1}+\\
2n_{-1}+2n_{0}
\end{array}
$ & $2{\Omega }\left[ \frac{
1}{2{\Omega }}\right] \left[ T\right] _{\frac{1}{2{\Omega }}}\left[
T+1\right] _{\frac{1}{2{\Omega }}}$ & $\left| n,T ,i\right\rangle $ \\ \hline
$u_{q}^{0}(2)$ &
$\begin{array}{c}
F_0,\ G_0,\\
\frac{N-2\Omega }{2};\ T _0
\end{array}$
& $i=n_{1}-n_{-1}$ &
$\begin{array}{c}
2{\Omega }\left[\frac{1}{2{\Omega }
}\right] \left[ \frac{2\Omega -2(n_1+n_{-1})}{2}\right]
_{\frac{1}{2{ \Omega }}} \\
\times \left[ \frac{2\Omega -2(n_1+n_{-1})}{2}+1 \right] _{
\frac{1}{2{\Omega }}}
\end{array}$
  & $\begin{array}{l}
\left| n_{1},n_{0},0\right) \\
\left| 0,n_{0},n_{-1}\right)
\end{array} $ \\ \hline
$u_{q}^{\pm }(2)$ &
$\begin{array}{c}
F_{\pm 1},\ G_{\pm 1},\\
\frac{N_{\pm 1}-\Omega }{2};\ N_{\mp 1}
\end{array}$
& $2n_{\mp 1}+n_{0}$ &
$\rho _{\pm }{\ \Omega }\left[
\frac{1}{{\Omega }}\right] \left[ \frac{{\Omega }{-}n_0}{2}
\right] _{\frac{1}{{\Omega }}}\left[ \frac{{\Omega }{-}n_0}{2}{
+1}\right] _{\frac{1}{{\Omega }}}$ & $\begin{array}{l}
\left| n_{1},0,n_{-1}\right) \\
\left| n_{1},1,n_{-1}\right)
\end{array}$ \\ \hline
\end{tabular}
\label{tab:limits}
\end{table}

The generalized model has the same symmetry properties as the single-level
realization of the theory \cite{fermRealSp4}. All formulae that are constructed
in terms of commutation relations of the single-level generators 
(like action of a
group generator on the basis states, Casimir invariants, eigenvalues, 
normalization
coefficients of the basis vectors) coincide at the algebraic level and have the
same form under the substitution $\Omega _j
\rightarrow
\Omega$. The three important reduction limits of the
$sp_q(4)$ algebra to $u_q(2)$ are summarized in Table \ref{tab:limits} with the
eigenvalues of the first and second-order Casimir invariants and the 
basis states
given for each limit.  Also, the
$q$-deformed symplectic algebra reverts back to the ``classical'' limit when
$q\rightarrow 1$.

For nuclear structure applications we use the set of the
commutation relations that is symmetric with respect to the exchange $
q\leftrightarrow q^{-1}$ \cite{fermRealSp4}. The $q$-coefficients, 
$\Psi _{lk}(N_{p})$,
obtained in \cite{fermRealSp4} for the following  commutation relations
\[
\fl [T_l,Y_{\pm k} ]=\pm \frac{Y_{\pm l \pm k}\Psi_{\pm 1}(N_{\pm
k})}{2[2]\sqrt{\Omega }}, \ l,k \ne 0; \ \ \ [F_l,G_{-k}
]=\frac{T_{l+k}\Psi_{|l|-|k|}(N_{l-k})}{2[2]\sqrt{\Omega }},\ l+k\ne 0,
\]
(where $Y=F\ (G)$ for the `$+$' (`$-$') case)  can be written in a 
compact way as
\begin{equation}
\Psi _{\pm 1}(N_{p})=
2\sqrt{\rho_+\rho_-}\left[ 2_{N_{p}\pm 1/2-\Omega 
}\right]_{\frac{1}{2\Omega}}=\left\{
\begin{array}{l}
\Psi _{l0}(N_{p}) \\
\Psi _{0k}(N_{p})
\end{array}
\right.,
\label{Psi}
\end{equation}
where we define $[2_{X}]_{
\frac{1}{2\Omega }}\equiv \frac{\left[ 2X\right] _{\frac{1}{2\Omega }}}{
\left[ X\right] _{\frac{1}{2\Omega }}}=q^{\frac{X}{2\Omega 
}}+q^{-\frac{X}{2\Omega
}}\stackrel{q\rightarrow 1}{\rightarrow }2$.

\section{Action of first and second-order operators on the basis vectors}

In addition to a generalization of the $sp_q(4)$ algebra to multi-$j$ shells,
an algebraic form for the action of the product of two or more 
generators of the
symplectic symmetry can be given. This allows one to calculate the
overlaps between states in a number preserving sequence, to build a 
$q$-deformed
second-order diagonal operator in terms of all the ten generators of 
$Sp_q(4)$, and
to obtain the matrix elements of a model Hamiltonian.

\subsection{The $su^T_q(2)$ limit}

In the $q$-deformed case, the commutators of the raising (lowering) $T_{\pm
}$ operator with the pair creation operators, $(F_{\mp })^{n_{\mp
1}}$ and $(F_0)^{n_0}$, which enter into the construction
of the basis states (\ref{qGencsF}), are
\begin{eqnarray}
\left[T_{\pm },(F_{\mp })^{n_{\mp
1}}\right]=F_0(F_{\mp })^{n_{\mp 1}-1}\frac{\sqrt{\rho_+\rho_-}}{\sqrt{\Omega
}[2]}[n_{\mp 1}]_{\frac{1}{2\Omega}}\left[ 2_{N_{\mp 1}+n_{\mp 1}-1/2-\Omega
}\right]_{\frac{1}{2\Omega}} \nonumber \\
\left[T_{\pm },(F_0)^{n_0}\right]_{\left( \frac{[2]}{2}\right)^{n_0}}=F_{\pm
}(F_0)^{n_0-1} \frac{1}{2\sqrt{\Omega
}}\sum_{p=0}^{n_0-1}\frac{[2]^p}{2^p} [2_{N_{\mp 
1}+n_0-1-p}]_{\frac{1}{2\Omega}}
\label{lrT}
\end{eqnarray}

With the use of (\ref{lrT}), the general formula for the action of 
the $k$-th order
product of $T_{\pm }$ on the lowest (highest)
weight basis state can be determined,
\begin{eqnarray}
T_{\pm }^k(F_{\mp })^{n_{\mp 1}}\left| 0\right>&=&\sum_{i=0}^{\lfloor 
k/2 \rfloor }
\left( \frac{\sqrt{\rho_+\rho_-}}{\sqrt{\Omega}[2]}
\left[ 2_{n_{\mp 1}-\frac{1}{2}-\Omega }\right]_{\frac{1}{2\Omega}} 
\right)^{k-i}
\frac{[n_{\mp 1}]_{\frac{1}{2\Omega}}!\ \theta(k,i)}{[n_{\mp 1}-k+i]_
{\frac{1}{2\Omega}}!} \nonumber \\
&& \times (F_{\pm })^{i}(F_0)^{k-2i}(F_{\mp })^{n_{\mp 1}-k+i}\left| 0\right>,
\label{kTonBasis}
\end{eqnarray}
where $k \le n_{\pm 1} $ and the functions in the sum are defined as
\begin{eqnarray}
\fl \theta(k,0)=1, \forall~ k \nonumber \\
\fl \theta(k,i)=\left\{
\begin{array}{l}
\frac{\theta(k-1,i)\left[ 2_{n_{\mp 1}-i-\frac{1}{2}-\Omega
}\right]_{\frac{1}{2\Omega}}}{\left[ 2_{n_{\mp 1}-\frac{1}{2}-\Omega
}\right]_{\frac{1}{2\Omega}}} +\frac{\theta(k-1,i-1)}{2\sqrt{\Omega
}}\sum_{p=0}^{k-2i}\frac{[2]^p}{2^p} 
[2_{k-2i-p}]_{\frac{1}{2\Omega}},\ i\le \lfloor k/2
\rfloor \\
0,\ i>\lfloor k/2 \rfloor
\end{array}
\right..
\end{eqnarray}
This implies that
$\theta(k, \frac{k}{2})=\frac{\theta(k-1,
\frac{k}{2} -1)}{\sqrt{\Omega }}$ when $k$ is even. Starting from the lowest
(highest) weight basis state the action of the $T_{\pm }$ operator 
(\ref{kTonBasis})
gives all the number preserving vectors with a definite maximum value 
of the $T$
quantum number. (Recall that for given $n$ and $i$, $T$ takes the values,
$T=\frac{\tilde{n}}{2},\ \frac{\tilde{n}}{2}-2,...,\ 2 \lceil 
\frac{i}{2} \rceil$,
where $\tilde{n}=\min\{n,4\Omega -n\}$.) The rest of the vectors with
lower $T$ values and the same
$(n,i)$ quantum numbers can be found as independent and orthogonal vectors to
those constructed in (\ref{kTonBasis}).

In nuclear systems, the $T_{\pm }$ generators represent the raising 
and lowering
isospin operators and as such they generate $\beta ^{\mp }$-decay 
transitions in an
isobaric sequence. It follows that formula (\ref{kTonBasis}) derived
above is used extensively in the calculation of the strength of these
transitions. Also, the construction of the isospin states
(\ref{kTonBasis}) allows one to compute overlaps with the pair states
(\ref{qGencsF}) and with the  eigenvectors of  a model
Hamiltonian.

\subsection{Action of products of two generators on the basis states}

We are also able to give an analytical form of the action of the
anticommutator $\{T_{+},\ T_{-}\}=T_+T_-+T_-T_+$ on the basis states
\begin{eqnarray}
\fl \{T_{+},T_{-}\}\left| n_{1},n_{0},n_{-1}\right) 
&=\frac{M^T_{-1,+2,-1}}{\Omega }
\left| n_{1}-1,n_{0}+2,n_{-1}-1\right)   \nonumber \\
\fl &+ \frac{M^T_{0,0,0}}{\Omega } \left| n_{1},n_{0},n_{-1}\right)
+\frac{M^T_{+1,-2,+1}}{2\Omega } \left| n_{1}+1,n_{0}-2,n_{-1}+1\right) ,
\label{syme}
\end{eqnarray}
where the coefficients $M^T_{n'_1,n'_0,n'_{-1}}$ (with 
$n'_1,n'_0,n'_{-1}$ indicating
the number of pairs of each kind added ($+$)/removed ($-$)) are 
$q$-deformed functions
of the pair numbers given in terms of the $n_{-1},n_0,n_{1}$ by
\begin{eqnarray}
M^T_{-1,+2,-1}&=&\frac{1}{4[2]^2}\{\Psi (n_{0},n_{1}-1)\Psi (n_{0}+1,n_{-1}-1)
\nonumber \\
&+&\Psi (n_{0},n_{-1}-1)\Psi (n_{0}+1,n_{1}-1)\} \nonumber \\
M^T_{0,0,0}&=&\frac{1}{4[2]}\{\Phi (n_{0}-1)\left(\Psi (n_{0}-1,n_{1})+\Psi
(n_{0}-1,n_{-1})\right)  \nonumber\\
&+&\Phi (n_{0})\left(\Psi (n_{0},n_{1}-1)+\Psi
(n_{0},n_{-1}-1)\right)\} \nonumber \\
M^T_{+1,-2,+1}&=&\Phi (n_{0}-1)\Phi (n_{0}-2),
\label{Ms}
\end{eqnarray}
where we define
\[
\fl \Phi (n_{0})=\sum_{k=0}^{n_0} 
\frac{\left[2\right]^k}{2^k}\left[2_{n_0-k}\right]_{\frac{1}{2\Omega }},
\]
\begin{eqnarray}
\fl \Psi (n_{0},n_{\pm 1})&=&\left[ n_{0}+2n_{\pm 1}+1\right] 
_{\frac{1}{2\Omega }
}-\left[ n_{0}-1\right] _{\frac{1}{2\Omega }}+\left[ n_{0}+2n_{\pm
1}+2-2\Omega \right] _{\frac{1}{2\Omega }}-\left[ n_{0}-2\Omega \right] _{
\frac{1}{2\Omega }} \nonumber \\
\fl &=& 2\sqrt{\rho_+\rho_-}\left[ n_{\pm }+1\right] _{\frac{1}{2\Omega
}}\left[2_{n_0+n_{\pm }+1/2-\Omega }\right]_{\frac{1}{2\Omega }}.
\label{Psi}
\end{eqnarray}
The second expression of $\Psi (n_{0},n_{\pm 1})$ (\ref{Psi}) allows the
non-diagonal term that scatters two identical particle pairs of 
opposite kinds into
two non-identical particle pairs (\ref{Ms}) to be rewritten as
\begin{eqnarray}
\fl M^T_{-1,+2,-1}&=\frac{\rho_+\rho_-}{[2]^2}\left[
n_1\right]_{\frac{1}{2\Omega}}\left[ n_{-1}\right]_{\frac{1}{2\Omega}} \times
\nonumber \\
\fl &\{ \left[2_{n_0+n_1-\Omega -\frac{1}{2}}
\right]_{\frac{1}{2\Omega }}\left[2_{n_0+n_{-1}-\Omega +\frac{1}{2}}
\right]_{\frac{1}{2\Omega }}+\left[2_{n_0+n_1-\Omega +\frac{1}{2}}
\right]_{\frac{1}{2\Omega }}\left[2_{n_0+n_{-1}-\Omega -\frac{1}{2}}
\right]_{\frac{1}{2\Omega }} \}.
\end{eqnarray}

In a similar way, the action of the second-order product, $F_0G_0$, 
on the basis
states yields the following non-diagonal term
\begin{eqnarray}
-\frac{1}{\Omega}M^P_{-1,+2,-1}&=&-\frac{1}{\Omega
}\tilde{n}_{1}\tilde{n}_{-1} \nonumber \\
&=&-\frac{\rho_+\rho_-}{[2]^2 \Omega }\left[
n_1\right]_{\frac{1}{2\Omega}}\left[ n_{-1}\right]_{\frac{1}{2\Omega}}
\left[2_{n_1-\Omega -\frac{1}{2}}
\right]_{\frac{1}{2\Omega }}\left[2_{n_{-1}-\Omega -\frac{1}{2}}
\right]_{\frac{1}{2\Omega }},
\label{Mp0}
\end{eqnarray}
and for $F_{+1}G_{+1}+F_{-1}G_{-1}$ it is
\begin{eqnarray}
-\frac{1}{\Omega }M^P_{+1,-2,+1}=-\frac{1}{\Omega }\frac{\sqrt{\rho _+ \rho
_-}}{\left[  2\right]}\sum_{k=1}^{n_{0}-1}
S_q(k),
\label{Mppm}
\end{eqnarray}
where  we define
$\tilde{n}_{\pm 1}\equiv \frac{1}{2[2]} ( \left[ 2n_{\pm 1}-1 \right]
_{\frac{1}{2\Omega }}+ \left[ 2n_{\pm 1}-2\Omega \right]
_{\frac{1}{2\Omega }}+ \left[ 2\Omega \right]
_{\frac{1}{2\Omega }}+1 )=\frac{1}{[2]}\sqrt{\rho_+ \rho_-}\left[ n_{\pm
1}\right] _{\frac{1}{2\Omega }}\left[ 2_{n_{\pm 1}-\Omega -1/2}\right]
_{\frac{1}{2\Omega }}$
$\stackrel{q\rightarrow 1}{\rightarrow }n_{\pm 1}$, and $S_q(k)\equiv
[2_{k-\Omega -1/2}]_{\frac{1}{2\Omega }} \sum_{i=0}^{k-1}\frac{\left[ 2\right]
^{i}}{2^{i}}[2_{k-1-i}]_{\frac{1}{2\Omega }}\stackrel{q\rightarrow
1}{\rightarrow }4k$ \cite{algPairing}. The diagonal elements, $M^P_{0,0,0}$, of
$F_0G_0$ and $F_{+1}G_{+1}+F_{-1}G_{-1}$ are discussed in detail in
\cite{algPairing}.

\subsection{Second-order operators}

The analytical relations (\ref{Ms})-(\ref{Mppm}) allow us to
find a $q$-deformed second-order operator, $O_2(sp_q(4))$, that is 
diagonal in the
$q$-deformed basis and that in the limit when
$q$ goes to one reverts to the second-order Casimir invariant of the
$sp(4)$ algebra \cite{GoLiSp,fermRealSp4},
\begin{eqnarray}
\fl O_2(sp_q(4)) =
\frac{\gamma _1}{2} (\{F_{+1},G_{+1}\}+\{F_{-1},G_{-1}\})
+\gamma _0\frac{C_2(su_q^0(2))}{\Omega } +\frac{C_2(su_q^T(2))}{\Omega },
\label{qC2}
\end{eqnarray}
where the $\gamma$-coefficients are $q$-functions of the pair numbers,
$\gamma _1=
\frac{M^T_{+1,-2,+1}}{2M^P_{+1,-2,+1}}\textstyle{\stackrel{q\rightarrow
1}{\rightarrow }}2$ and $\gamma _0=\frac{
\left[2_{n_0+n_1-\Omega -\frac{1}{2}}
\right]_\frac{1}{2\Omega}\left[2_{n_0+n_{-1}-\Omega +\frac{1}{2}}
\right]_\frac{1}{2\Omega}+\left[2_{n_0+n_1-\Omega +\frac{1}{2}}
\right]_\frac{1}{2\Omega}\left[2_{n_0+n_{-1}-\Omega -\frac{1}{2}}
\right]_\frac{1}{2\Omega} }{2
\left[2_{n_1-\Omega -\frac{1}{2}}
\right]_\frac{1}{2\Omega}\left[2_{n_{-1}-\Omega -\frac{1}{2}}
\right]_\frac{1}{2\Omega}}\stackrel{q\rightarrow  1}{\rightarrow }1$. 
The Casimir invariants in (\ref{qC2}) are
$ C_2(su_q^T(2))={\Omega }(\{T_{+},T_{-}\}{+}\left[ \frac{1}{{\Omega }}\right]
\left[ T_{0}\right] _{\frac{1}{2{\Omega }}}^{2})$ and
$ C_2(su_q^0(2))={\Omega }(\{F_{0},G_{0}\}{+}\left[ \frac{1}{{\Omega }}\right]
\left[ \frac{N}{2}-\Omega \right] _{\frac{1}{2{\Omega }}}^{2}) $ 
\cite{fermRealSp4}.

The second-order operator can be written in terms of the Casimir
operators of all four limits, $\{+,-,0,T\}$, as
\begin{eqnarray}
\fl O_2(sp_q(4)) = \sum_{k={+,-,0,T}}\gamma _k \frac{C_2(su_q^k(2))}{\Omega }
-\frac{\gamma _1}{2} \left[\frac{2}{\Omega } \right]\{ \rho_+\left[
\frac{N_1-\Omega }{2}\right]^2_\frac{1}{\Omega } + \rho_-\left[
\frac{N_{-1}-\Omega }{2}\right]^2_\frac{1}{\Omega }
\}
\label{qC2casSU2}
\end{eqnarray}
where $\gamma _{\pm} \equiv \gamma _1,\ \gamma _T \equiv 1$ and
$C_2(su_q^{\pm }(2))=
\frac{{\Omega }}{2}(\{F_{\pm 1},G_{\pm 1}\}{+}\rho _{\pm }\left[
\frac{2}{{\Omega }}\right] \left[ \frac{N_{\pm 1}-\Omega}{2}\right] 
_\frac{1}{{\Omega }
}^2)$ \cite{fermRealSp4}. Its eigenvalue in the basis set (\ref{qGencsF}) (see
Table
\ref{tab:limits} and (\ref{Ms})-(\ref{Mppm})) is
\begin{eqnarray}
\fl \left<O_2(sp_q(4)) \right>&=&
\gamma _1 (\rho_+ +\rho_-) \left[
\frac{1}{{\Omega }}\right] \left[ \frac{{\Omega }{-}n_0}{2}
\right] _{\frac{1}{{\Omega }}}\left[ \frac{{\Omega }{-}n_0}{2}{
+1}\right] _{\frac{1}{{\Omega }}} \nonumber \\
&&-\frac{\gamma _1}{2} \left[\frac{2}{\Omega } \right] \{ \rho_+\left[
\frac{2n_++n_0-\Omega }{2}\right]^2_\frac{1}{\Omega } + \rho_-\left[
\frac{2n_-+n_0-\Omega }{2}\right]^2_\frac{1}{\Omega } \}
\nonumber \\
&&+2\gamma _0\left[\frac{1}{2{\Omega }
}\right] \left[ \frac{2\Omega -2(n_1+n_{-1})}{2}\right]
_{\frac{1}{2{
\Omega }}}\left[ \frac{2\Omega -2(n_1+n_{-1})}{2}+1 \right] _{
\frac{1}{2{\Omega }}}
\nonumber \\
&&+\frac{M^T_{0,0,0}}{\Omega }+
\left[\frac{1}{\Omega } \right]\left[n_1-n_{-1}\right]^2_\frac{1}{2\Omega
}\ \ \ \stackrel{q\rightarrow  1}{\rightarrow }\Omega +3.
\end{eqnarray}

The second-order operator (\ref{qC2}) is a Casimir invariant only in 
the non-deformed
limit of the theory. In that limit its eigenvalue $(\Omega +3)$ 
labels the $Sp(4)$
representations. While an explicit form for the second-order Casimir 
operator of
$sp_q(4)$ for other $q$-deformed schemes can be given \cite{aach}, 
this is not the
case here because the scheme includes, 
by construction \cite{fermRealSp4}, a
dependence on the shell structure which is suitable for nuclear 
physics applications.
Nevertheless, the importance of the second-order operator (\ref{qC2}) in the
$q$-deformed case is obvious. It is an operator that consists of 
number preserving
products of all ten $q$-deformed group generators, and the pair basis states
(\ref{qGencsF}), which span the entire space for a given $Sp_q(4)$ 
representation,
are its eigenvectors. Its zeroth-order approximation commutes with 
the generators
of the $q$-deformed symplectic symmetry, which means that only the higher-order
terms introduce a dependence on the quantum numbers that label the states.
It also gives a direct relation between the expectation values of the 
second-order
products of the generators that build $O_2(sp_q(4))$.

The analytical formulae, which were derived above, are also used 
for finding the
matrix elements of the interaction in a system with symplectic 
dynamical symmetry.
The model Hamiltonian \cite{algPairing} is another second-order 
operator that is
expressed in terms of the generators of the $Sp_q(4)$ group
\cite{KleinMarshalek,fermRealSp4}:
\begin{eqnarray}
\fl H_{q} =&-\bar{\epsilon}^{q}N
-G_{q}F_{0}G_{0}-F_{q}(F_{+1}G_{+1}+F_{-1}G_{-1})
-\frac{E_{q}}{2\Omega }(C_2(su_q^T(2))-\Omega \left[ \frac{N}{2\Omega }\right])
\nonumber \\
\fl &-C_{q}2\Omega \left[ \frac{1}{\Omega }
\right] (\left[ \frac{N}{2}-\Omega \right] ^2_{\frac{1}{2\Omega }}
- \left[ \Omega \right]^2 _{\frac{1}{2\Omega }})
-(D_{q}-\frac{E_{q}}{2\Omega })\Omega
\left[
\frac{1}{\Omega }\right] \left[ T _{0}\right] ^2_{\frac{1}{2\Omega }},
\label{qH}
\end{eqnarray}
where $\epsilon ^{q}=\bar{\epsilon}^{q}+(\frac{1}{2}-2\Omega 
)C_{q}+\frac{D_{q}}{4}>0$
is the Fermi level of the system, $G_{q},\ F_{q},\ E_{q},\ C_{q}$ and $D_{q}$
are constant interaction strength parameters. The classical 
Hamiltonian, $H_{cl}$, is
obtained in the limit $q\rightarrow 1$. For a nuclear system with $N_+$ valence
protons and $N_-$ valence neutrons, the interaction represents 
proton-neutron and
identical-particle isovector pairing (with
$G_q\geq 0$ and
$F_q\geq 0$ strength parameters), a symmetry term ($E_q$), a diagonal 
proton-neutron
isoscalar force ($E_q$ and $C_q$) 
\cite{HasegawaKanekoPRC59,KanekoHasegawaPRC60} and an
isospin breaking term ($D_q \neq \frac{E_q}{2\Omega }$). The quantum 
extension of the
$sp(4)$ algebra introduces higher-order interactions and accounts for 
non-linear
effects. The $q$-deformed model can be applied to multi-$j$ major shells,
like $1f_{5/2}2p_{1/2}2p_{3/2}1g_{9/2}$, and is exactly solvable. The
second-order diagonal operator
$O_2(sp_q(4))$  (\ref{qC2}) sets a linear
dependence between the
$q$-deformed pairing and symmetry energies, which allows for a 
reduction of the number
of the phenomenological parameters in the Hamiltonian (\ref{qH}).

\subsection{Boson approximation}

Although the fermion generalization allows many $j$-orbitals to be 
considered, the
dimension of the space should not be allowed to grow too large 
because the effect of
the deformation diminishes as the size of the model space grows. For 
example, in the
case of very large $\Omega $ the anticommutation relation of the fermions
(\ref{qfcr}) reduces to the simpler form
$\{\alpha _{j,\sigma ,m},\alpha _{j,\sigma ,m^{\prime }}^{\dagger }\}_{q^{\pm
1}}=\delta _{m,m^{\prime }}$ and all
$q$-brackets $[X]$ or $[X]_{\frac{1}{(2)\Omega }}$ go to
$X$  when X is not a function of $\Omega $. In this limit the 
pair-operators obey
boson commutations and a boson approximation is achieved. The model Hamiltonian
(\ref{qH}) is diagonal in the pair states and does not include 
scattering between
different kinds of bosons. In the limit of large $\Omega \gg
\{\frac{N}{2},N_{\pm 1}\}$, the action of the operators $A\cdot B$ 
($F\cdot G$) on the
basis states counts the number of different kinds of bosons:
$\left<F_0G_0\right>$ counts the number of non-identical particle 
pairs and remains
non-deformed; $\left<F_{\pm }G_{\pm }\right>$ counts the number of identical
particle pairs and only scales its non-deformed analog by a factor of 
$\frac{1+q^{\pm
1}}{2}$. As another direct consequence of the dependence of the 
deformation on the
space dimension is that in this large $\Omega $ limit the direct product of the
single-$j$ quantum symplectic algebras holds, 
$sp_q(4)=\bigoplus_{j}sp_q^{(j)}(4)$,
as for the non-deformed case for all $\Omega $.

\section{Conclusion}

In this article we introduced a multi-shell extension of the quantum $sp_q(4)$
algebra. While in the non-deformed case this is a direct sum of the single-$j$
symplectic $sp_q^{(j)}(4)$ algebras, in the deformed case a direct 
sum result is
only achieved in a boson approximation of the theory that is 
applicable in the large
space limit,  $\Omega \gg \{\frac{N}{2},N_{\pm 1}\}$. The dependence of the
deformation on the dimensionality of the space makes the 
generalization unique and
non-trivial.

We also derived an analytical solution for the action of the 
$q$-deformed raising
and lowering operators on the basis states. This, in turn, allows one 
to calculate
the overlap between number preserving states. It also makes the
construction of $q$-deformed basis vectors with a definite isospin value
possible and allows one
to calculate $\beta$-decay transition probabilities between these states.

We were also able to obtained formulae for computing the action of 
the product of
two generators of the $Sp_q(4)$ group on the basis states.  From this 
we found a
$q$-deformed second-order operator in the group generators that is 
diagonal in the
basis set with its zeroth-order approximation commuting with all the 
$Sp_q(4)$ generators. The results can also be used to provide for an 
exact solutions
of a $q$-deformed model Hamiltonian.

\vskip .5cm This work was supported by the US National Science 
Foundation, Grant
Numbers 9970769 and 0140300. The authors thank Dr. Vesselin G. 
Gueorguiev for use of
his MATHEMATICA programs for non-commutative algebras.

\end{document}